This Article has been accepted for publication in IEEE CiSE

# Uncovering Scientific Software Sustainability through Community Engagement and Software Quality Metrics

Sharif Ahmed, *University of Central Arkansas, Conway, AR, USA*

Addi Malviya Thakur, *University of Tennessee Knoxville; Oak Ridge National Laboratory, TN, USA*

Gregory R. Watson, *Oak Ridge National Laboratory, Oak Ridge, TN, USA*

Nasir U. Eisty, *University of Tennessee, Knoxville, TN, USA*

*Abstract—Scientific open-source software (Sci-OSS) projects are critical for advancing research, yet sustaining these projects long-term remains a major challenge. This paper explores the sustainability of Sci-OSS hosted on GitHub, focusing on two factors drawn from stewardship organizations: community engagement and software quality. We map sustainability to repository metrics from the literature and mined data from ten prominent Sci-OSS projects. A multimodal analysis of these projects led us to a novel visualization technique, providing a robust way to display both current and evolving software metrics over time, replacing multiple traditional visualizations with one. Additionally, our statistical analysis shows that even similar-domain projects sustain themselves differently. Natural language analysis supports claims from the literature, highlighting that project-specific feedback plays a key role in maintaining software quality. Our visualization and analysis methods offer researchers, funders, and developers key insights into long-term software sustainability.*

## Introduction

Open-source software (OSS) has become central to technological advancement, with much of today's software developed collaboratively by diverse contributors. However, such efforts sometimes struggle to sustain thriving, innovative projects due to factors like disengagement, conflicts, and technical or management challenges [1]. Researchers have examined both successful and abandoned projects to understand what makes them sustainable, often using the term *software sustainability*, which refers to the long-term viability of the software, not its environmental sustainability. Approaches to measuring sustainability include ecological analogies alignment with the United Nations' Sustainable Development Goals and socially or economically sustainable processes [2]. While recent work has explored sustainability by quantifying software energy consumption [3], this alone has proven insufficient for assessing software quality. Thus, defining meaningful metrics to measure OSS sustainability remains a challenge, as noted in other studies [4], [5].

Scientific open-source software (Sci-OSS) comprises tools and applications designed to address scientific challenges, thereby advancing research across various fields, including physics, biology, and engineering. Unlike general OSS, Sci-OSS is designed for the specific needs of scientific research, often incorporating unique algorithms and data analysis frameworks. Sustainability is critical for Sci-OSS because research projects often require tools that remain accessible and reliable over long periods [6].

Sustainable Sci-OSS fosters high-quality, adaptable, and reproducible software that can evolve with technological advancements, supporting ongoing collaboration among researchers. Moreover, sustaining these tools maximizes the return on investment, often funded publicly, by ensuring the software continues to benefit the scientific community without constant redevelopment. Therefore, Sci-OSS sustainability underpins the accuracy, reliability, and longevity essential for impactful, reproducible scientific research.

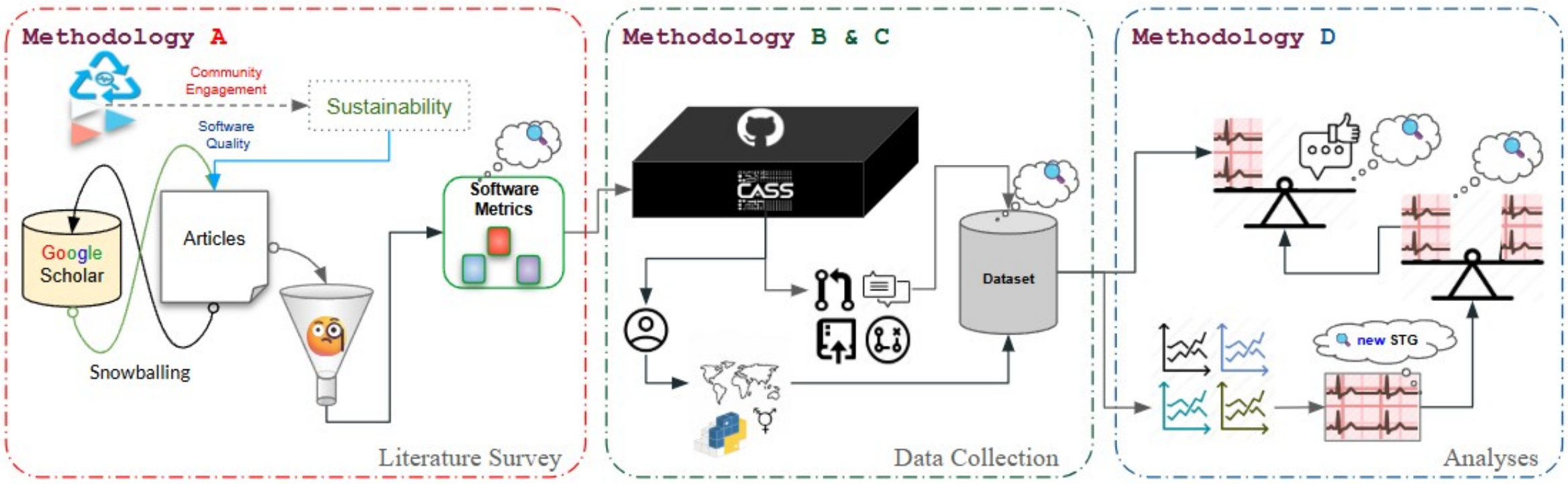

FIGURE 1: Study methodology: identified sustainability-related software metrics from the literature (left), extracted these metrics from 10 open-source scientific projects (middle), and analyzed the data by developing a new visualization technique (right).

Clearly, there is a need for sustainability measurement for Sci-OSS projects that transform science and regulate technology, as well as those that involve grants or other funding agencies. **Therefore, our goal for this study is to investigate the sustainability of Sci-OSS. To do so, we identified two major aspects from the missions and visions of software sustainability stewardship organizations, namely: community engagement and software quality.** These aspects are the first step to quantifying the sustainability of Sci-OSS. Due to the diverse definitions and perspectives we discovered in the literature, deriving a single formula for measuring sustainability proved unlikely. At this point, our next step was to find repository-related metrics (e.g., cyclomatic complexity, issue labels, response time, response type) that describe community engagement and software quality of Sci-OSS. These metrics from the social coding platforms and source codes can collectively measure the sustainability of Sci-OSS. We refer to these metrics as *repository metrics* for the rest of our paper. To address these challenges, we examine how community engagement and software quality metrics can reveal the long-term sustainability of Sci-OSS projects.

We pose the following research questions to guide our study.

**RQ1: What role do community engagement and software quality metrics play in sustaining Sci-OSS projects?**

**RQ2: How do the repository metrics related to software sustainability differ among diverse Sci-OSS projects?**

**RQ3: Does the utility of code review feedback in Sci-OSS projects differ?**

To sum up, the contributions of our study are as follows:

- A dataset mined from ten well-known GitHub Sci-OSSs
- A novel sustainability visualization technique
- Identification of patterns of Sci-OSS sustainment through community engagement and software quality metrics
- Statistical and linguistic comparisons from the mined data

## Methodology

This section details our methods to understand the Sci-OSS sustainability and FIGURE 1 shows an overview.

### A. Factors and Metrics for Sustainability

OSS is key to the success of many scientific and commercial software. Most OSS projects are hosted on a small number of social coding platforms such as GitHub or GitLab. These platforms typically provide access to a wide range of metrics related to software development and software maintenance practices (e.g., commit history, code patches, issue tracking, developers' discussion, review feedback and related artifacts.) Often, these metrics are used as an approximate measure of the software quality assurance [7]. In this paper, we aim to find the software metrics from social coding platforms to quantify the sustainability of Sci-OSS.

***Leading factors*** According to the Better Scientific Software (BSSw) [F1], "*Sustainable software means that an existing product remains viable in the future such that it makes sense to continue using, adapting, and expanding its capabilities instead of replacing it with another new or existing product.*" For sustainable science, stewardship organizations such as IDEAS[F2], CHAOSS[F3], ReSA[F4], CORSA[F5], and CASS[F6] have been supporting Sci-OSS projects. Their objectives guide us to aspects that intersect sustainability with software metrics from repository and hosting platform data. From this, we have identified two major aspects of the missions and visions of these organizations: **community engagement** and **software quality**.

***Metrics*** After identifying the leading factors, we reviewed literature that discuss the factors, software sustainability-related topics, or both. We performed forward and backward snowballing using Google Scholar. Next, we obtained features and factors from the relevant papers and filtered out non-retrievable or non-computable factors and features. Finally, we established a list of software metrics derived from sustainability related lead factors, *community engagement* and *software quality*. TABLE 1 shows the final list of repository metrics, their aspects, and rationale.

## B. Data Collection

After extracting the sustainability-related software metrics from the literature, we investigated public GitHub scientific software repositories. We explored the US Department of Energy (DOE)-sponsored Software Stewardship Organizations (SSOs) and randomly selected 10 projects that are open and active on GitHub. Here, we excluded the projects that are mirrored on GitHub from social coding or other platforms because the mirrored repositories entirely or partially lack many of the GitHub-provided metrics like issues and pull requests (PRs). In this paper, we anonymize the project names to prevent bias or unintended endorsement or criticism. The software metrics shown in TABLE 1 are grouped into the two leading factors community engagement and software quality. They are also highlighted as primary or secondary metrics. The metrics data directly available or computable from the GitHub REST API responses or PyDriller are named 'primary metrics' in our paper. We obtain these metrics as outlined below.

1) We collected the GitHub issues, PRs, and issue/review comments data using GitHub APIs.
2) We chose PyDriller, a Python framework, to mine commits instead of GitHub REST APIs due to API usage restrictions and the enormous number of commits. By cloning the selected GitHub projects, we used PyDriller to mine commit messages, code patches, and other metrics.
3) We processed the mined primary data and obtained secondary metrics data as described in the next subsection.

## C. Data Processing

After collecting the primary metrics data, we computed the secondary metrics data using existing tools and research artifacts. We detail our methods to compute the secondary metrics below.

***Repository Metrics*** We obtained primary metrics data directly from our mined data. These included an aggregated enumeration of issues, PRs, issue comments, pull request comments, commits, labels, new labels, emoji reactions on issue/PR, and commits authored [F7] [F8] [F9]. We also enumerated the merged, closed, or open PRs and then open or closed issues [F18]. Since software activity is most frequently studied at a monthly granularity, we aggregated our metrics by month. TABLE 1 shows these metric names preceded with #.

***Temporal Metrics*** The GitHub API provided time information on PRs, issue, commit, and comment creation; PR, issue, commit, and comment modification; PR and issue closure; and PR merge. We converted all the times into UTC zones. For closed PR/issue enumeration, we selected the calendar month from the GitHub *closed at* timestamp values. For PR and issue closure duration metrics, we subtracted the creation-timestamp from the closure-timestamp.

***Model-computed Metrics*** To evaluate sentiment, usefulness, toxicity, and readability of PRs and issue comments, we selected a state-of-the-art sentiment

[F1] https://bssw.io/items/what-is-software-sustainability
[F2] https://ideas-productivity.org/overview/impact.html
[F3] https://chaoss.community
[F4] https://www.researchsoft.org
[F5] https://corsa.center
[F6] https://cass.community

[F7] https://doi.org/10.1109/TSE.2014.2349496
[F8] https://doi.org/10.1109/SANER.2017.7884682
[F9] https://doi.org/10.1145/3368089.3409746

TABLE 1: Software Metrics

Here, : Community Engagement : Software Quality aspects;
Metrics starting with * indicates *primary* otherwise *secondary*

| A. | Factors | Source | Rationale | Metrics |
|---|---|---|---|---|
| | community attention | issue, PR | The attention (too rapid/ slow response) regulates the engagement of newcomers F7. Average response time, number of responses, and number of closed issues were proposed as indicators for sustained activity F28 | issue/ PR response time<br>*# issue/ PR comments<br>* #emoji reactions |
| | peer performance | issue, commit | Minimum productivity (issues/month) of the peers F7 | * #issues created |
| | coding & review activity | commit PR | More contributions receive more feedback F8. PyPI developers perceive it as a main indicator for sustained activities F28 | * # review comments<br>*# PR created<br>* #commits |
| | altruism | issue/ PR-comments | Selfless contribution to improve & share software for the benefit of the community F12 | sentiment<br>usefulness |
| | community hostility | issue/PR-comments | Developers leave OSS for hostile interactions F13 | toxicity |
| | entry barriers | issue | Issues supporting newcomers indicates less barriers and more new participation F9 | labeled as #good-first-issue or #help-wanted |
| | readability | issues | Newcomers choose the more understandable issues F14 | PR/ issue readability |
| | user base | users | PyPI developers perceive that the number of issue reporters is an effective measure for sustained activities F28 | *#issue reporters<br>*# PR creators<br>*# commit authors |
| | spatial/ cultural distances | commit, profile | Spatial/ cultural distances are key factors for FLOSS projects F20 | location coverage<br>user-type variation<br>association heterogeneity |
| | labels | issues PR | Issue labels help developers and reviewers to navigate and choose appropriate issues | * #labels<br># new labels |
| | bug fixing activity | issues | Bug fixing is required to ensure the software remains trustworthy, and implementing new features is required to ensure the software remains relevant. | * # issues closed |
| | feedback | PR-comments | The reviews need to be useful F11 | usefulness |
| | open vs closed issues | issues | Ensure software usability for future F18 | * #closed-issues/ #open-issues |
| | cyclomatic complexity | commit | It is associated with maintainability of a project F15 | cyclomatic complexity |
| | deduplication | issues | Avoid rework and support newcomers F14 | is deduplicated |
| | gender, minority etc. | profile | Gender ratio increases novelty and impact in scientific ideation F18 | gender ratio |
| | maintainability | commit | Fine grained technical debt F16 | unit interfacing<br>method size |
| | closure length | issues<br>PR | Indicates the complexity or pace of the project | PR closure duration<br>issue closure duration |
| | authorship | commit | Company domination negatively relates with the survival probability of OSS projects F21 | affiliation heterogeneity |

analyzer [F10], useful/not-useful predictor [F11] [F12], toxicity detector [F13]and readability tool [F14] respectively. We measured the maintainability of projects using PyDriller over mined commits, which provided scores for delta-maintainability-model's (dmm) cyclomatic complexity, unit interfacing, and method size [F15] [F16].

***Community Belonging and Engagement (CBE) Metrics*** TABLE 1 shows that *affiliation, user type, association, gender,* and *location* secondary metrics are related to measuring CBE. We approximated the developers' gender from their names using a gender-guessing technique [F17] to measure gender ratio, an important metric for community engagement [F18] [F19] [F20] [F21]. We also took the developers' location data from their GitHub profiles to measure geo-coverage. For maximum coverage, we normalized the location information using GeoPy [F22] geocoder and took only the country names. Lastly, we opted for Shannon's Diversity Index to calculate the variation of given software metrics categorical values:

$$H' = -\sum_{i=1}^{R} p_i \ln(p_i)$$

Here, $p_i$: proportion of belonging to the $i^{th}$ category.

[F10]https://doi.org/10.1109/ASE.2017.8115623
[F11]http://doi.org/10.1007/s10664-025-10617-1
[F12]https://doi.org/10.1080/10864415.2002.11044241
[F13]https://doi.org/10.1145/3583562
[F14]https://doi.org/10.1145/3544902.3546236
[F15]https://doi.org/10.1109/TSE.1976.233837
[F16]https://doi.org/10.1109/TechDebt.2019.00030
[F17]https://pypi.org/project/gender-guesser
[F18] https://doi.org/10.1371/journal.pcbi.1010627
[F19] https://doi.org/10.7717/peerj-cs.156
[F20] https://doi.org/10.1145/2487294.2487330
[F21] https://doi.org/10.1145/3540250.3549117
[F22]https://pypi.org/project/geopy

## D. Data Analysis

We performed exploratory data analysis on collected and processed data as follows.

***Visual Analysis*** To understand the sustainability of scientific software through our compiled metrics, we employed data visualization techniques including line, histogram, box, and violin plots in order to see trends, distributions, and characteristics. For better visualization, we used a *log-scaled metric value* on the y-axis and the *year-month* of the scientific projects on the x-axis. Since the compiled software metrics are many and comprise different types of data with various ranges, traditional visualization techniques were difficult to apply. This led us to develop an alternative, robust, visualization technique which we have called the ***S**oftware sus**T**ainability **G**raph (**STG**)* and which is detailed in the following Section. The STG helped us look into the progression of the metrics over time more clearly and answer our research question **RQ1**.

***Statistical Analysis*** Based on our intuition and observations from exploratory visual analysis , we attempted to deepen our understanding using statistical hypothesis testing. We selected the non-parametric *t*-test functionality provided by SciPy [F23] for the Wilcoxon signed-rank test to contrast the snapshots of our selected Sci-OSSs. To assess the level of statistical significance, we checked the p-values from the *t*-test, employing threshold benchmarks of 0.05 (5%), 0.01 (1%), and 0.001 (0.1%) with Holm-Bonferroni [8] adjustment. Next, we checked the magnitude of the significant differences using Cliff's Delta for our paired comparisons. For cross-project comparison, we took the mean values of raw STG components for each project over the entire length. These statistical analyses collectively answer **RQ2.**

***Conversation Analysis*** With the intuition from visual analysis and evidence from statistical analysis, we next examined the discussions between the code author and code reviewer. This analysis complements the visual and numerical insights with developers' communication. We utilized a conversation analysis toolkit [9] to perform the analysis in the context of code review discussion as follows.

1) We prepared the ConvoKit corpus by transforming Utterance, Conversation, and Speaker instances from the PR discussions.
2) We added the usefulness of the PR comments [10], [11] that are manipulated in our original dataset (see Data Processing) as meta-data.
3) We preprocessed the corpus with ConvoKit TextCleaner.
4) We trained ConvoKit's FightingWords transformer, a conversation analyzer, with our prepared and cleaned corpus. We trained the transformer for each project separately to obtain model outputs separately. In our case, a positive log-odds ratio indicated that a word is more associated with the useful class, while a negative one is with the not-useful class.
5) After getting the generated summary and scores from ConvoKit, we examined how utterances of useful and not-useful comments differ across the projects.

This analysis between useful and not-useful code review comments helps to answer **RQ3**.

# Software Sustainability Graph (STG)

## Foundation

In our work, we identified 2 leading factors from sustainability stewardship efforts, then 10 factors for each leading factor, and then 31 unique metrics while answering **RQ1**. Many of these metrics are applied to different sources. For example, sentiment can be analyzed for both issue and PR comments. Traditional visualization of each metric produces too much information that can not be easily interpreted to determine the sustainability profile of a software project. FIGURE 2a also shows that visualizing multiple metrics from a project or multiple projects for a metric can be difficult to comprehend. Though **small multiples** [F24] seems like an alternative, a recent work [12] discovered in their eye-tracking analyses that the utility of small multiples declines in accuracy with increasing frames due to human cognitive capacity limits. An electrocardiogram (ECG) presents a wealth of information in a very compact form that retains a predictive capability, we adopt the ECG concept to visualize repository metrics data of Sci-OSS. FIGURE 2c shows our proposed visualization technique, **S**oftware Sus**T**ainability **G**raph (STG), which has 18 STG-leads (analogous to ECG leads) and 46 STG-components from software metrics (see TABLE 1) that come from community engagement and software quality repository metrics that lead us to quantify sustainability of a software project.

[F23] https://scipy.org/

[F24] https://en.wikipedia.org/wiki/Small_multiple

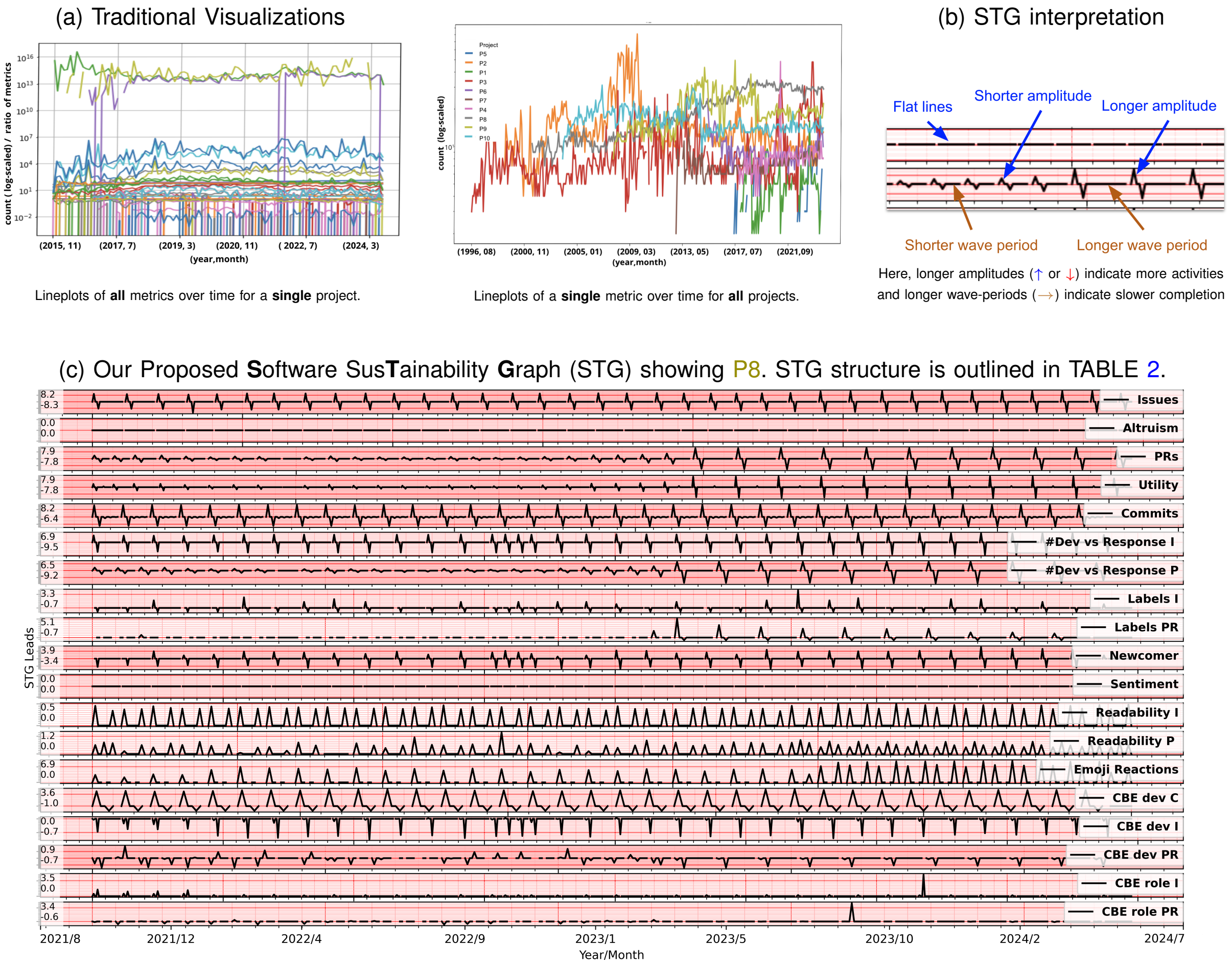

FIGURE 2: Metrics Visualized Using Traditional and Our Proposed Visualization

## STG Structure

We have structured STG (FIGURE 2c) where

- STG leads are based on factors and STG components are metrics from TABLE 1.
- TABLE 2 shows STG leads and their components.
- Position and direction of the STG component are also shown in the table with ↑, ↓, and → markers: ↑ for the positive y-axis, ↓ for the negative y-axis, and → for the positive x-axis.
- The ↑ causes crests and ↓ causes troughs in our STG.
- Here, STG components on the x-axis (→) are time durations.
- Three STG leads, namely *commits*, *emoji-reactions*, and *CBE Developer C*, do not have any time durations (→ components).

## STG Interpretation

- The amplitudes in the STG lead signals are mostly in log scale, larger amplitudes mean higher values (See FIGURE 2b).
- The period or interval between the peaks and/or troughs (→) is log-scaled time durations in seconds.
- The disconnection of components separates the aggregated components (e.g., day, week, month, or year).
- The wave period of each cycle demonstrates the productivity; the shorter, the better.
- The consecutive peak and trough are related to software metrics, such as issues created (peak ↑) followed by issues closed (trough↓), to show their contrasts.
- The flat lines indicate no values or zero values.

TABLE 2: **S**oftware Sus**T**ainability **G**raph (STG) Structure

Here, ↑: positive y-axis, ↓: negative y-axis, →: the positive x-axis (Time duration/ response time on → is always in seconds and log-scaled), and CBE: Community Belonging and Engagement.

| STG lead | STG components |
|---|---|
| Issues | ↑ log-scaled number of issues created, ↓ log-scaled number of issues closed, → issue closure duration |
| Altruism | ↑ ratio of useful issue comments, ↓ ratio of toxic issue comments, → issue closure duration |
| PRs | ↑ log-scaled number of PRs created, ↓ log-scaled number of PRs closed, → PR closure duration |
| Utility | ↑ ratio of useful review comments, ↓ ratio of toxic review comments, → PR closure duration |
| Commits | ↑ log-scaled total commits, ↓ log-scaled total authors, ↓ unit interfacing, ↓ cyclomatic complexity, ↓ method size |
| Developer Response I | ↑ log-scaled number of issue reporters, ↓ log-scaled number of issue comments, → average issue response time |
| Developer Response PR | ↑ log-scaled number of PR creators, ↓ log-scaled number of review comments, → average PR response time |
| Labels-I | ↑ log-scaled number of new issue-labels, ↓ log-scaled number of total issue-labels, → average issue response time |
| Labels-PR | ↑ log-scaled number of new PR-labels, ↓ total log-scaled number of PR-labels, → average PR response time |
| Newcomer Support | ↑ log-scaled number of newcomer issues, ↓ log-scaled number of deduplicated issues, → issue response time |
| Sentiment (support) | ↑ median issue comments' sentiment, ↓ median review comments' sentiment , → average issue response time |
| Readability I/PR | ↑ issue/PR comments readability, ↓ issue/PR body readability, → average issue/PR response time |
| Emoji Reactions | ↑ total reactions on issues, ↓ total reactions on issue-comments ↑ total reactions on review-comments (all log-scaled) |
| CBE developer C | ↑ affiliation heterogeneity in commits, ↓ number of parent commits |
| CBE developer I | ↑ gender ratio in issues, ↓ location coverage in issues, → average issue response time |
| CBE developer PR | ↑ gender ratio in PR, ↓ location coverage in PR, → average PR response time |
| CBE roles I | ↑ association heterogeneity in issues ↓ user type variation in issues, → average issue response time |
| CBE roles PR | ↑ association heterogeneity in PR, ↓user type variation in PR, → average PR response time |

TABLE 3: Overview of selected projects (anonymized)

| | | | Contributors | | Uses | | | | | | Open | | Closed | |
|---|---|---|---|---|---|---|---|---|---|---|---|---|---|---|
| **Name** | **created** | **language** | **unknown** | **known** | **starred** | **forks** | **followers** | **labels** | **size** | **commits** | **issues** | **PR** | **issues** | **PR** |
| P1 | 2018-11 | Python | 34 | 26 | 267 | 60 | 18 | 14 | 44697 | 2775 | 21 | 1 | 196 | 71 |
| P2 | 2015-11 | C++ | 659 | 218 | 1174 | 559 | 116 | 147 | 817241 | 101853 | 395 | 72 | 12775 | 7859 |
| P3 | 2018-11 | C | 124 | 45 | 640 | 182 | 31 | 10 | 223645 | 12254 | 189 | 27 | 911 | 566 |
| P4 | 2017-12 | C | 43 | 39 | 187 | 45 | 25 | 44 | 19474 | 5008 | 46 | 13 | 1577 | 1325 |
| P5 | 2017-06 | Fortran | 17 | 10 | 23 | 16 | 14 | 12 | 2053 | 249 | 7 | 1 | 72 | 61 |
| P6 | 2015-04 | C++ | 209 | 139 | 1793 | 412 | 83 | 32 | 33158 | 13141 | 481 | 82 | 6625 | 4111 |
| P7 | 2012-11 | C++ | 104 | 57 | 657 | 146 | 37 | 33 | 372273 | 22474 | 460 | 15 | 1247 | 191 |
| P8 | 2016-12 | LLVM | 6218 | 370 | 26665 | 10926 | 593 | 353 | 3049433 | 503435 | 24509 | 2646 | 71957 | 21858 |
| P9 | 2016-10 | C | 167 | 103 | 518 | 276 | 42 | 38 | 89219 | 19816 | 242 | 65 | 6775 | 3552 |
| P10 | 2014-09 | C | 389 | 270 | 2065 | 845 | 118 | 57 | 179659 | 34377 | 776 | 97 | 11864 | 8978 |

## Results and Discussion

### Data Overview

***Projects*** TABLE 3 shows the list of the selected projects[F25] and snapshots of their GitHub metrics at the time of data mining. All of the projects had active contributions and user engagement. Metrics such as stargazers, forks, and followers are potentially useful for studying sustainability but are not directly retrievable using the GitHub API for specific past periods. We are only reporting them in this subsection to provide an idea of the studied projects.

***Dataset*** This study mined 141,687 GitHub issues, 184,536 issue-comments, 52,172 pull-requests (PRs), 164,708 PR-comments, and examined 706,384 commits from the selected projects along with their associated computed metric values described in Methodology-C.

[F25] We have anonymized the projects to respect the developers' contributions while focusing on the broader analysis.

## RQ1: Role of Metrics

For **RQ1**, we identified 31 unique metrics through sustainability leading factors: community engagement and software quality (TABLE 1). It will also help software projects, software sustainability researchers, and sustainability stewardship organizations better understand the landscape of software sustainability.

Our proposed **S**oftware Sus**T**ainability **G**raph (STG), described in earlier Section and shown in FIGURE 2c, provides an overview within a compact-sized image. Due to space limitations, we show the last 12 months' snapshot of selected projects in FIGURE 3. Our artifact [F26] also includes snapshots of the recent 1 year, 3 years, 5 years, 10 years, and the entire project duration. Even though the projects *P1-10* vary in terms of user base, software category, age, or stage (TABLE 3), we found the sustainability of the selected projects from our STG visualization and its interpretation as follows.

- The **issue** lead, which shows issues created and closed in a month seems to be the most rhythmic compared to other leads for all projects. Some projects take time and close their issues in the following month(s), for example, *P5*. A possible scenario would be when an issue was reported on the last day/week of a month and resolved on the following day/week of a new month. The longer issue closure duration (wave period) within the projects is mostly followed by low or no issue creation (crest) and/or issue closure (trough), e.g., *P5, P1*, and *P7*.
- The **PR** lead overall resembles the **issue** lead, but this lead exhibits slower and irregular amplitude growth. Possibly, many of the contributors are not able to participate in PRs due to access restrictions, lack of expertise, etc. Additionally, some problems take longer time to solve or review than other problems.
- The **altruism** and **utility** leads show that all the projects have zero or negligible amounts of toxic comments (troughs in the STG). Flat lines indicate either no comments or comments deemed not useful. Interestingly, we see the disappearance of **altruism** values for *P8* before half-life, *P2* after five years, and *P9* after two years, but they have the continuous appearance of useful review comments for **utility** lead. This would seem to imply that projects might invest time in understanding why reviews are not proving to be useful, which could in turn help to improve the overall sustainability of the software.
- The **utility** lead shows more than 50% of the review comments are useful on average for all of the projects. We find flat lines for the **utility** lead is mostly due to the absence of review comments as we compare **utility** lead with the *developer-response PR* lead.
- The **developer vs. response** leads for both issues and PRs demonstrate different patterns. The response time for issues in more recent months is less than in the previous months, but the response time for PRs is more periodic. However, we find the *P2* and *P10* projects have only one denser pattern (higher productivity) around the middle of their regular patterns. Additionally, the *P8* and *P9* projects show regular denser patterns (more productive) at regular intervals. A possible reason could be the availability of maintainers. It indicates that participants are getting attention not immediately but later. These periodic concentrated efforts clear up the communication debts and help sustain the overall project flow.
- The **readability** lead informs us that they are mostly regular, and pull-requests (PRs) are more readable than the issues for all of the projects.
- The **newcomer-support** lead shows that newcomer-supported issue labeling and issue de-duplication are irregular and arrhythmic. Also, the denser response time found for issues' **developer response** lead is only visible for *P8* project here. Lastly, we find that the sentiment scores are mostly flat for all the projects all the time. These components of STG can help projects recognize a need to improve onboarding processes.
- The most irregular lead we find is the **labels** lead. For most of the projects, new labels are introduced for issues, and pull-requests(PRs) are more sudden than the overall assignment of labels. However, we find a high number of labels created for *P8* project in recent times and high usage of labels for *P10* project in the past. The usage of labels is diverse; for example, some projects use them to cluster focusing target users, some use them for knowledge areas or skills, and some use them for grouping difficulty. The variability might be the cause of the non-deterministic patterns.
- The non-verbal **emoji reactions** lead shows that *P6* and *P10* projects have a rhythmic bell-curve on issue-emojis. The *P8* and *P9* have an

[F26] https://github.com/sharif509/STG-artifact

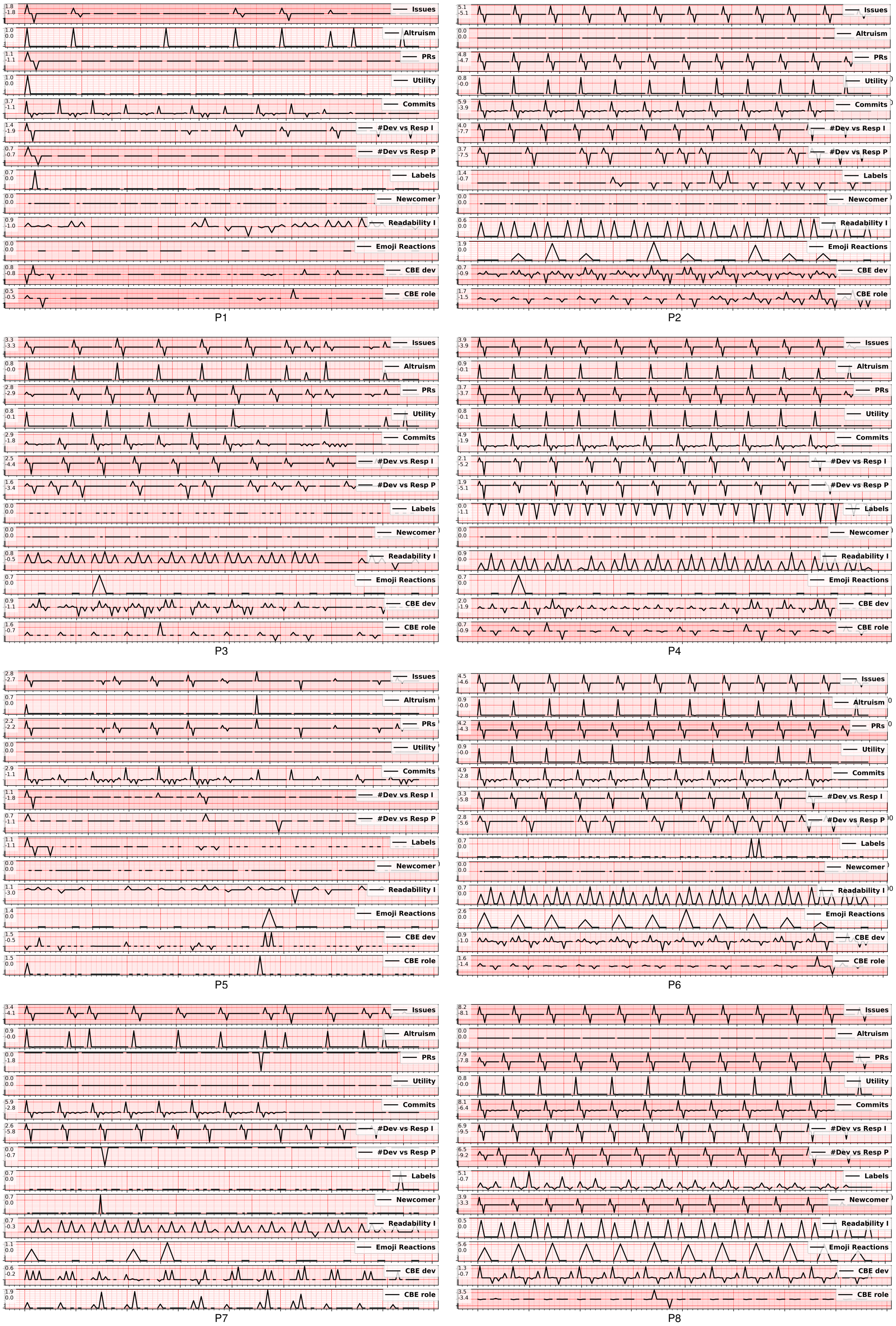

FIGURE 3: Sustainability Graph of Studied Projects: P1-8

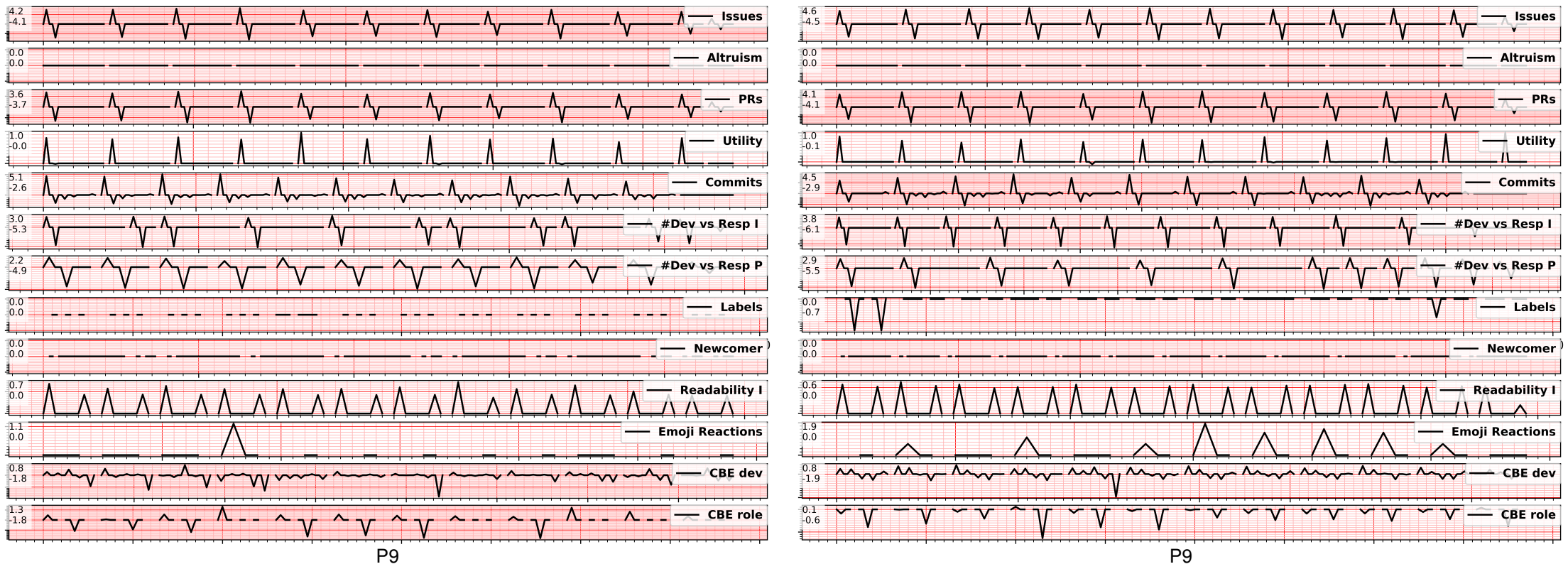

FIGURE 3: Sustainability Graph of Studied Projects: P9-10

increase in emoji reactions on issues over time. The *P4, P7* and *P3* show a decrease in emoji reactions on issues. For reactions on issues or review comments, we observe a sawtooth pattern for *P4* project, a very skewed pattern for *P8* only in recent months, and an irregular or scattered pattern for the remaining three projects: P1, P2, and P5.

- The **commit** lead shows a reverse wedge progression for *P2, P7,* and *P8*. Regarding *P8*, the number of commit-authors significantly increases throughout the course of time. The rest of the projects: *P1, P3, P4, P5, P6, P9,* and *P10* show either uniform or irregular patterns.
- The **maintainability** in commit lead shows that *cyclomatic complexity* has chaotic patterns for *P3, P5, P9,* & *P10* projects and *P6* seems almost always complex. Regarding *number of parent commits*, only *P8* has zero parent commits, i.e., it has linear commit history. The *P8* project has consistently large-sized methods. Then *P2* and *P6* show mostly larger methods. The rest of the projects have mixed-sized methods. In the case of *unit interfacing*, *P6* and *P8* projects tend to be tightly coupled. By understanding which of these applies to a particular project's code base, it should be possible for stakeholders to direct resources to the areas that would most benefit from sustainability investment.
- The **CBE role heterogeneity** leads reveal that the *P5, P1, P7,* and *P8* projects have notable association and user-type variation for PRs whereas *P10* has significant variation for issues. The rest of the projects have proportional variation for both issues and PRs.
- The **CBE developer heterogeneity** leads report that affiliation variation rises and falls for most of the projects. Only *P8* and *P6* have an overall increase. *P8* and *P6* also show positive changes in several community engagement and software quality metrics, which backs the findings that reducing company domination can have a net benefit on the project. Second, gender ratio patterns are more prominent for issues than pull requests. Third, geo-location coverage patterns are similar to gender ratio with slightly increased amplitudes. Another observation is that the number of countries is much higher than the number of gender categories, which may explain why our analysis finds that geo-coverages have higher amplitudes than gender-ratio scores.

**To summarize, the repository metrics contribute to identifying and measuring community engagement and software quality of the Sci-OSS.**

## RQ2: Comparison across Diverse Sci-OSSs

Based on observations from the results of our visual analysis in **RQ1**, we find that *sentiment* component is not contributing to the projects' sustainability. So, we omit *sentiment* and take the rest of the unique components from our Software Sustainability Graph (STG) for every project. Next, we obtain the statistical differences among the projects from our comparison setup (Methodology-D).

TABLE 4 shows the cross-project comparisons. It shows that our paired *t*-test fails to reject the null hypothesis (p-value $\geq$ 0.05) for most of the projects when compared pairwise. The effect size and p-values are presented in boldface. Importantly, P1 shows significant differences with all other projects (p-value $\leq$ 0.001). Though P5 and P7 do not show differences

TABLE 4: Cross Project Comparison of STG Components

| ↓ Project → | P5 | P2 | P1 | P3 | P6 | P7 | P4 | P8 | P9 | P10 |
|---|---|---|---|---|---|---|---|---|---|---|
| P5 | | $0.33^{*}$ | $0.06^{-}$ | $0.15^{-}$ | $0.27^{**}$ | $0.08^{-}$ | $0.27^{-}$ | $0.38^{***}$ | $0.25^{*}$ | $0.31^{*}$ |
| P2 | $-0.33^{*}$ | | $-0.32^{***}$ | $-0.21^{-}$ | $-0.08^{-}$ | $-0.28^{**}$ | $-0.13^{-}$ | $0.09^{-}$ | $-0.11^{-}$ | $-0.02^{-}$ |
| P1 | $-0.06^{-}$ | $0.32^{***}$ | | $0.13^{***}$ | $0.25^{***}$ | $0.04^{**}$ | $0.24^{***}$ | $0.36^{***}$ | $0.22^{***}$ | $0.3^{***}$ |
| P3 | $-0.15^{-}$ | $0.21^{-}$ | $-0.13^{***}$ | | $0.13^{-}$ | $-0.08^{-}$ | $0.09^{-}$ | $0.27^{-}$ | $0.1^{-}$ | $0.17^{-}$ |
| P6 | $-0.27^{**}$ | $0.08^{-}$ | $-0.25^{***}$ | $-0.13^{-}$ | | $-0.21^{-}$ | $-0.04^{-}$ | $0.17^{-}$ | $-0.03^{-}$ | $0.07^{-}$ |
| P7 | $-0.08^{-}$ | $0.28^{**}$ | $-0.04^{**}$ | $0.08^{-}$ | $0.21^{-}$ | | $0.19^{-}$ | $0.35^{-}$ | $0.19^{*}$ | $0.26^{**}$ |
| P4 | $-0.27^{-}$ | $0.13^{-}$ | $-0.24^{***}$ | $-0.09^{-}$ | $0.04^{-}$ | $-0.19^{-}$ | | $0.19^{-}$ | $0.0^{-}$ | $0.09^{-}$ |
| P8 | $-0.38^{***}$ | $-0.09^{-}$ | $-0.36^{***}$ | $-0.27^{-}$ | $-0.17^{-}$ | $-0.35^{-}$ | $-0.19^{-}$ | | $-0.19^{-}$ | $-0.1^{-}$ |
| P9 | $-0.25^{*}$ | $0.11^{-}$ | $-0.22^{***}$ | $-0.1^{-}$ | $0.03^{-}$ | $-0.19^{*}$ | $-0.0^{-}$ | $0.19^{-}$ | | $0.09^{-}$ |
| P10 | $-0.31^{*}$ | $0.02^{-}$ | $-0.3^{***}$ | $-0.17^{-}$ | $-0.07^{-}$ | $-0.26^{**}$ | $-0.09^{-}$ | $0.1^{-}$ | $-0.09^{-}$ | |

Here, the effect-size scores are from *Cliff's Delta* and their superscripts $*$: p-value$<$0.05 ,$**$: p-value$<$0.01 , and $***$: p-value$<$0.001 from the Wilcoxon signed-rank test with Holm-Bonferroni adjustment.

themselves, each of them shares statistical differences with P10, P7, and P2. Additionally, P5 shows difference with P6 and P8. The P2, P3, and P4 belong to the same scientific domain and did not show any differences, as expected.

In the human life cycle, people of the same age may have different heart conditions. Even healthy people of different ages may have good heart conditions but different functioning patterns. Thus, ECG reports of these people may have mixed patterns. So do the STGs of Sci-OSSs.

**These results appear to indicate that there is no clear distinction between projects of various sizes and states or from diverse domains.**

## RQ3: Utility of Code Review Feedback

We obtain very high-level linguistic characteristics from artifact/Figs/RQ3[F26]. It shows which words or phrases distinguish the utility (i.e., useful or not useful) of code review discussion within our selected projects except for *P5* project. We found that the *P5* project has only 19 comments, and all of them are predicted as useful. In contrast, the ConvoKit tool requires utterances from more than one category. We also see the log-odds ratio ranges differ from one project to another.

We found that project-specific words are useful for the *P8*, *P6*, and *P4* projects as their project names have appeared as significant utterances. The words such as *line, should, number, suggestion, you, see* have appeared as useful for more than one project. The commonly found not-useful uttered words are *this, this pr, change, ok, sounds good, good, better, we, same, why*, etc. We also see that the projects from different domains have common words for both useful and not-useful code review comments. After our high-level observation, we manually looked into the z-scored words from ConvoKit's conversation analysis. artifact/Figs/RQ3 annotates the top 10 words for both classes and leaves the in-between data points without annotation, so as expected, we see that common English words are in useful and not-useful classes.

Overall, the list of useful words appears to be indicative, such as referring to a line of code or the need to do something. On the other hand, non-useful words mostly resemble first-person personal pronouns, typical affirmations, and other linguistic fillers. The take-away from this is that while Sci-OSS introduces many new words, it still resembles much of the software-specific jargon used in software development.

**This analysis uncovered that projects that differ in their sustainability capability still utter common things related to sustaining the software quality.**

## Threats to Validity

- The derived metrics rely on data from varied sources (literature, repositories, GitHub API), which may introduce unknown biases or inconsistencies.
- Gender and location information are often imprecise due to anonymized, encrypted, or improperly filled GitHub profiles, limiting demographic-based analyses.
- Findings may not generalize beyond the studied scientific OSS domains (applied math, programming systems, machine learning), and may not reflect patterns in broader or non-scientific open-source ecosystems.
- Statistical choices such as log-scaling and the use of non-parametric *t*-tests help mitigate some analytical biases, but visual and statistical interpretations may still be constrained by data transformation and scale choices.

## Related Work

We find several approaches that define and measure software sustainability. The majority of these approaches are longitudinal studies that try to forecast or predict factors like developers' leaving ratios, core/ peripheral developers, defect removal/ bug fixing ratios[F27], contributors' activities[F28], graduation[F29], or to check these factors' longevity. We argue that predicting a sustainability-related factor is too fine-grained for predicting holistic sustainability.

For holistic sustainability, we found multiple approaches, such as analogizing software lifecycles with ecological processes[F30], investigating how open-source software contributes to the United Nations' 17 Sustainable Development Goals[F31], modeling software product for transparent and sustainable society or economy[F32]. Although these contributions are valuable for software research, these approaches cannot measure the sustainability of a given code and software development metrics data. Recently, Pathania et al.[F33] quantified software sustainability by calculating the energy consumption of the software's source code. They observed that their energy efficiency based sustainability scores conflict with maintaining software quality.

## Conclusion

We presented a multimodal analysis (visual, statistical, and conversational) of Sci-OSS software sustainability, identifying key factors, community engagement, and software quality, reflected through repository metrics. We introduced the STG, aggregating these metrics to visualize sustainability trends. Future work includes expanding metric coverage, refining STG with new features, and integrating these tools into practical workflows to support sustainable Sci-OSS development. Our results provide a foundation for integrating sustainability metrics directly into research software evaluation pipelines, enabling researchers and institutions to better assess, compare, and support long-term software sustainability.

[F27] https://doi.org/10.1016/j.is.2014.10.005
[F28] https://doi.org/10.1145/3236024.3236062
[F29] https://doi.org/10.1145/3468264.3468563
[F30] https://doi.org/10.48550/arXiv.1309.1810
[F31] https://www.linuxfoundation.org
[F32] https://doi.org/10.1145/3643655.3643883
[F33] https://doi.org/10.1109/ASE56229.2023.00204

## Acknowledgment & Data Availability

This manuscript has been authored by UT-Battelle, LLC, under contract DE-AC05-00OR22725 with the US Department of Energy. The publisher acknowledges the US government license to provide public access under the DOE Public Access Plan[F34].

**Artifact:** https://github.com/sharif509/STG-artifact.

[F34] https://energy.gov/downloads/doe-public-access-plan

**Sharif Ahmed** is an Assistant Professor in the Department of CSE, University of Central Arkansas, Conway, AR, USA. Contact him at sahmed@uca.edu

**Addi Malviya Thakur** is Software Engineering Group Leader in the Computer Science and Mathematics Division at Oak Ridge National Laboratory, TN, USA. Contact her at amalviya@vols.utk.edu or malviyaa@ornl.gov

**Gregory R. Watson** is Application Engineering Group Leader in the Computer Science and Mathematics Division at Oak Ridge National Laboratory, Oak Ridge, TN, USA. Contact him at watsongr@ornl.gov

**Nasir U. Eisty** is an Assistant Professor in the Department of EECS, University of Tennessee, Knoxville, TN, USA. Contact him at neisty@utk.edu